\documentclass[a4paper,11pt]{article}
\usepackage[T1]{fontenc}
\usepackage[utf8]{inputenc}
\usepackage{amsfonts}
\usepackage{amssymb}
\usepackage{amsmath}
\usepackage{lmodern}
\usepackage{graphicx}
\usepackage{hyperref}
\usepackage[usenames,dvipsnames]{color}
\usepackage{graphicx}

\usepackage{float}
\newtheorem{Theorem}{Theorem}
\newtheorem{Lemma}{Lemma}
\newtheorem{Definition}{Definition}

\newtheorem{Proposition}{Proposition}

\newtheorem{Condition}{Condition}
\newtheorem{Conditions}{Geometric Conditions}

\renewcommand{\le}{\leqslant}
\renewcommand{\leq}{\leqslant}
\renewcommand{\ge}{\geqslant}

\title{Generalised hyperbolicity in spacetimes with string-like singularities}
\author{Yafet Sanchez Sanchez\footnote{E-mail:Y.SanchezSanchez@soton.ac.uk} \\ 
and James A. Vickers\footnote{E-mail:J.A.Vickers@soton.ac.uk} \\ Mathematical Sciences and STAG Research Centre, \\ University of Southampton,\\ Southampton,\\ SO17 1BJ } 

\begin{document}

\maketitle
\begin{abstract} 
 In this paper we present well-posedness results
  for $H^{1}$ solutions of the wave equation for spacetimes that
  contain string-like singularities. These results extend a framework
  in which one characterises gravitational singularities as
  obstruction to the dynamics of test fields rather than point
  particles. In particular, we discuss spacetimes with cosmic
  strings.
\end{abstract}

\begin{section}{Introduction} 

  The identification of the gravitational field with the spacetime
  background in General Relativity makes the domain of the solution of
  Einstein's equations not known a priori. This requires that one
  considers local solutions and then looks for suitable extensions.
  One of the issues that needs to be considered when extending the
  metric is the regularity of the metric. Since the Penrose and
  Hawking singularity theorems \cite{hawking} the standard definition
  of a singularity has been in terms of geodesic incompleteness which
  intuitively can be thought of as an obstruction to the continuation
  of the world line of a free-falling observer.  Such a definition
  requires the metric to be at least $C^{1,1}$ in order to guarantee
  the existence and uniqueness of geodesics.  Furthermore, this
  regularity is the threshold where rough metrics and smooth metrics
  share the same causal structure \cite{c11,ch}.

  However, if we consider Einstein's field equations simply as a
  hyperbolic evolution system, solutions can be obtained with the
  metric in Sobolev spaces which are compatible with regularity below
  $C^{1,1}$ \cite {kato}.  Moreover, it has recently been shown that local well-posedness
  follows from having enough control over the $L^{2}$ norm of the curvature
  on the spatial foliation and the radius of injectivity \cite{l2}. From this point of view, the
  relevant condition to ensure the well-posedness of the evolution
  equations is not a pointwise condition on the curvature, but rather
  an $L^2$ type condition on the metric and its derivatives. In this
  context, an alternative is to consider a singularity as an
  obstruction to the evolution of a test-field rather than as an
  obstruction to the evolution of a particle, along a causal
  geodesic. This point of view was called generalised hyperbolicity
  by Clarke \cite{generalised}, and involves regarding certain
  traditional singularities as interior points in a spacetime with low
  regularity and then proving local well-posedness of the wave
  equation in the rough extension. The closely related concept of
  wave-regularity was introduced by Ishibashi and Hosoya \cite{ih}
  in their study of the wave equation in static singular spacetimes
  with timelike singularities. They used this term to define the
  well-posedness of the wave equation in the sense that unique
  solutions exist and there is no freedom in the boundary conditions
  one imposes on the singularity. A full discussion on this and
  related concepts can be found in \cite{ys, hypersurface}. 

  In previous work we looked at the concept of
  {\emph{wave-regularity}} for curve-integrable spacetimes \cite{ys}
  and spacetimes with singular hypersurfaces \cite{hypersurface}. In
  this paper we extend these results to spacetimes with string
  like singularities. These are timelike singularities of
  co-dimension 2. As explained below, a natural condition is to
  require the solutions to lie in the Sobolev space $H^1$ and for
  clarity we now state precisely what we mean by $H^{1}$-wave
  regularity:
  \begin{Definition} A point $p$ in $(M,g_{ab})$ is {\it $H^{1}$
      wave regular} if there is a region,
    $\Sigma_{(0,T]}=\Sigma\times (0,T]$, where $\Sigma$ is an open
    bounded region of an $n$-dimensional manifold containing $p$, such
    that the initial value problem on the initial hypersurface
    $\Sigma_{0}$ for the wave equation $\square_{g}u=f$ on
    $\Sigma_{(0,T]}$ is locally well posed in the following sense:
\begin{itemize}
     \item There exists a solution in the function space $H^{1}(\Sigma_{(0,T]})$
     \item The solution is unique in the function space $H^{1}(\Sigma_{(0,T]})$
     \item The solutions in the space $H^{1}(\Sigma_{(0,T]})$ is stable with respect to initial data.
     \end{itemize} 
A point $p$ is {\it weakly wave regular} if it only
     satisfies the first two conditions. A {\it strongly wave
       regular spacetime} is defined to be one such that every point
     $p$ in ${({\cal{M}}, g_{ab})}$ is strongly wave regular.
   \end{Definition}

Notice that, if a solution $u$ is
   in $H^1$ and the components of the metric are bounded,
   then the energy momentum satisfies $T^{ab}[u]\in
   L^{1}(\Sigma_{(0,T]})$ and can be defined as a
   distribution. Therefore, for locally bounded metrics,
   $H^{1}$-wave regularity implies the integrability of the energy
   momentum tensor of the solutions. Moreover, canonical quantisation
   schemes of free scalar fields require the existence of symplectic
   structures formed by products of the solution and its derivatives
   \cite{kay}. In general, these structures will be ill-defined for
   solutions with less regularity. This is particularly problematic
   if one expects to be able to take into account the quantum
   behaviour of the field. Additionally, if there do not exist $H^1$
solutions to the scalar wave equation there will not exist $H^1$
solutions to the linearised Einstein equations.

   In this paper, we provide tools to establish the existence and
   uniqueness of solutions of the wave equation with
   $H^{1}$-regularity in singular spacetimes. We also give examples of how
   $H^{1}$-wave regularity can be applied to a number of physically
   important scenarios. In \S 2 we prove a general theorem for wave
   equations in rough backgrounds with metrics that satisfy certain
   conditions (see geometric conditions \ref{con} in \S 2.1) which are
   satisfied by cosmic string type singularities for example. The
   basic proof of the theorem follows the method of Evans \cite[\S 7.2]{evans}
  and uses Galerkin approximation methods
   together with energy estimates for the wave operator.  The method
   of proof is different from that used in \cite{hypersurface} and
   reflects real differences in the type of singularity under
   consideration. For hypersurface singularities we were able to
   obtain energy estimates for the first order system and its adjoint,
   and use these to obtain existence using the Hahn-Banach
   theorem. For the string type singularities under consideration
   here, the special form of the metric means that we have good energy
   estimates for the wave operator but not for its adjoint in $H^{1}$. On the
   other hand, the time derivatives of the metric coefficients are
   well-behaved which is crucial to the use of a Galerkin
   approximation. This allows us to prove existence, uniqueness and
   stability of weak solutions with the required regularity. However
   the results differ from those in \cite{evans} in that we explicitly
   lower the differentiability (although see \cite{hunter}),
   generalise the results to curved spacetimes with a special emphasis
   on the $n+1$ decomposition of spacetime and use a
     different method of proof to establish uniqueness and stability
     which allows the result to be generalised if one works with more
     general gauges. Furthermore, we show that under the geometric conditions
   \ref{con} and the hypothesis of lemma \ref{l2} the energy momentum is not only integrable in spacetime,
   but can also be defined distributionally on any constant time
   hypersurface. In \S3 we discuss how our theorems apply to a large
   class of spacetimes with cosmic string type singularities and show
   that, from this perspective, cosmic string singularities should not
   be regarded as strong gravitational singularities, even though the
   curvature is not in $L^{\infty}_{loc}$, nor even in some cases in
   $L^{1}_{loc}$.  Finally at the end of the section, we discuss the
   relationship of our approach to other work and the importance of
   low regularity solutions for any discussion of the Strong Cosmic
   Censorship Conjecture.
   
\end{section}

{\bf{Notation.}} When considering the details of the function spaces
$L^{2},H^{1}$ we first write the domain and then the measure
considered, e.g., $L^{2}(\Sigma,\nu_{h})$. When the measure is the one
associated to the volume form $dx^{n}$, (or $dx^{n+1}$ when the domain
is the whole spacetime) we will omit the measure and just write
$L^{2}(\Sigma)$. The time dependence of functions will be explicitly
stated at the beginning of the theorems. However, the time dependence
will not be made explicit in calculations and estimates if there is no
risk of confusion. We also denote the derivative of a function $u$
with respect $t$ by $u_{t}$ and $u_{i}$ if it is with respect to the other
$x^i$-coordinates. When a function $d(t)$ depends only on time we
denote the derivative by $\dot{d}(t)$. Additionally to avoid
cumbersome notation we will not always explicitly use $\sum$ to denote
a sum and we use instead Einstein's summation notation, with roman
letters $a,b \dots$ etc used for summations over $0 \cdots n$ and $i,j \dots$ etc used for summations over $1 \cdots n$.  
We take the signature of the metric to be
$(+,-,-,...,-)$.

\begin{section}{The main theorem}

\subsection{The general setting} 

Let $\Sigma_{(0,T]}=\Sigma\times (0,T]$ be a $n+1$- dimensional domain
equipped with a Lorentzian metric $g_{ab}$ where $\Sigma$ is an open
bounded region of a $n$- dimensional manifold. Now using a $n+1$
decomposition of spacetime the line element of the metric may be
written in the form:
\begin{equation}
\label{generalmetric}
  ds^{2}= +N^{2}dt^{2}-\gamma_{ij}(dx^{i}+\beta^{i}dt)(dx^{j}+\beta^{j}dt)
  \end{equation}
where $N$ is the lapse function, $\beta^{i}$ is the shift  and $\gamma_{ij}$ is the induced metric on $\Sigma$.

The class of metrics we are going to consider requires that there is a
foliation of the domain $\Sigma_{(0,T]}$ and suitable coordinates
$(t,x^{i})$ such that

\begin{Conditions}\label{con}.
\begin{enumerate}
  \item $\gamma^{ij}\in C^{1}([0,T], L^{\infty}(\Sigma))$
  \item The volume form given by $\sqrt{\gamma}$,  for the induced metric $\gamma_{ij}$ is bounded from below by a positive real number, i.e., $|\sqrt{\gamma}|>\eta$ for $\eta\in \mathbb{R}^{+}$
    \item The lapse function $N$ can be chosen as $N=\sqrt{\gamma}$  
    \item The shift can be chosen in such a way that $\beta^{i}=0$
  \item There exist a constant $\theta>0$ such that
  $$\sum^{n}_{i,j=0}\gamma^{ij}{\gamma}\xi_{i}\xi_{j}\ge\theta|\xi|^{2}$$
  for all $(t,x)\in \Sigma_{(0,T]}, \xi\in\mathbb{R}^{n}$
  
\end{enumerate}
\end{Conditions}
Condition 3. on the lapse function can be weakened to require only that it is a bounded function with a positive lower bound i.e., 
\begin{equation*}
\begin{split}
 \emph{$3'.$} \mbox{ The lapse function } N  \mbox{ is } C^{1}((0,T], L^{\infty}(\Sigma)) \mbox{ and }  |N|>\omega \mbox{ for } \omega\in \mathbb{R}^{+} 
 \end{split}
\end{equation*}
However this is at the expense of adding linear terms in time and
to avoid undue complications in the formulae we do not pursue this
here. Note however the method of proof in \S \ref{prov} has been modified
from that in Evans \cite{evans} to allow for this possibility. 

We want to obtain weak solutions to the following initial/boundary
problem for the wave equation:
\begin{align}\label{pde3}
  \square_{g}u&=f \mbox{ in ${\Sigma_{(0,T]}}$}\\
  u&=0 \mbox{ on $\partial\Sigma\times [0,T]$}\\
 u(0,x)&=u_{0} \mbox{ on $\Sigma_{0}=\Sigma\times \{t=0\}$} \\
  u_{t}(0,x)&=h \mbox{ on $\Sigma_{0}=\Sigma\times \{t=0\}$}
\end{align} 
 where $f:\Sigma_{(0,T]}\rightarrow\mathbb{R}$ is a
given source and
$u_{0}:\Sigma\rightarrow\mathbb{R},h:\Sigma\rightarrow\mathbb{R}$ are
given initial conditions.

For a metric with a general $n+1$ splitting given by \eqref{generalmetric} the wave operator is given by:
\begin{equation}
\begin{split}
  \square_{g}u=\frac{1}{N\sqrt{\gamma}}\left(\partial_{t}\left(N\sqrt{\gamma}\frac{1}{N^{2}}\partial_{t}u\right)\right)\\
  +\frac{1}{N\sqrt{\gamma}}\left(\partial_{t}\left(N\sqrt{\gamma}\frac{\beta^{i}}{N^{2}}\partial_{i}u\right)+\partial_{j}\left(N\sqrt{\gamma}\frac{\beta^{j}}{N^{2}}\partial_{t}u\right)\right)\\
  -\frac{1}{N\sqrt{\gamma}}\partial_{i}\left(N\sqrt{\gamma}(\gamma^{ij}-\frac{\beta^{i}\beta^{j}}{N^{2}})\partial_{j}u\right)
 \end{split}
\end{equation}
%\frac{1}{\sqrt{-g}}\partial_{\mu}\left(\sqrt{-g}g^{\mu\nu}\partial_{\nu}u\right)\\

Taking into account the geometric conditions \ref{con} we obtain
\begin{eqnarray}
    \square_{g}u=\frac{u_{tt}}{{\gamma}}-\frac{Lu}{{\gamma}}\label{weak11}
  \end{eqnarray}

where $-L$ is an elliptic operator in divergence form given by:

\begin{equation}
  -Lu=-(\gamma^{ij}{\gamma}u_{j})_{i}
  \end{equation}
Notice that the geometric conditions imply that  $L$ is a uniformly elliptic operator.

We can associate to the operator $-L$ the bilinear form given by:

\begin{equation}
  B[u,v;t]:=\int_{\Sigma}\gamma^{ij}(t,x){\gamma(t,x)} u_{i}v_{j} dx^{n}
\end{equation}

\begin{Definition}\label{def}
We say a function:

$$u\in L^{2}(0,T;H^{1}_{0}(\Sigma)), \mbox{ with } u_{t}\in L^{2}(0,T;L^{2}(\Sigma)), u_{tt}\in L^{2}(0,T;H^{-1}(\Sigma))$$
is a local weak solution of the hyperbolic initial/boundary problem (\ref{pde3}) provided that locally:

\begin{enumerate}
\label{weakcs}
  \item For each $v\in L^{2}(0,T;H^{1}_{0}(\Sigma))$,
  \begin{equation}
  \begin{split}
    \int^{T}_{0}<u_{tt},v>dt+\int^{T}_{0}B[u,v;t]dt=(f,v)_{L^{2}(\Sigma_{(0,T]},\mu_{g})}\end{split}
  \end{equation} where $\mu_{g}=\sqrt{-g}d^{n+1}x={\gamma}d^{n+1}x$ and $<\cdot,\cdot>$ denotes the dual pairing between the $H^{-1}(\Sigma)$ and $H^{1}_{0}(\Sigma)$ Sobolev spaces. 
  \item $u(0,x)=u_{0}(x), u_{t}(0,x)=h(x)$ \label{weak2} where $u_{0}\in H^{1}_{0}(\Sigma_{0})$ and $h\in L^{2}(\Sigma_{0})$ 
\end{enumerate}

\end{Definition}

We motivate definition \ref{def} by the
following calculation. For the moment assume the metric and the
solution are smooth and that $\square_{g}u=f$. Multiplying by  an element $v\in
L^{2}(0,T;H^{1}_{0}(\Sigma))$ and integrating we obtain:

\begin{eqnarray}
 \nonumber 
\int_{\Sigma_{(0,T]}}f v \mu_{g} &=& \int_{\Sigma_{(0,T]}}(\square_{g}u) v \mu_{g} \\\nonumber
&=&\int_{\Sigma_{(0,T]}}\left(\frac{1}{{\gamma}}u_{tt}-\frac{1}{{\gamma}}Lu\right)v {\gamma} d^{n+1}x\\\nonumber
  &=&\int^{T}_{0}\int_{\Sigma}\left(u_{tt}-Lu\right)v d^{n}xdt\nonumber\\
  &=&\int^{T}_{0}\int_{\Sigma}\left(u_{tt}v\right)d^{n}xdt-\int^{T}_{0}\int_{\Sigma}(\gamma^{ij}\gamma u_{j})_{i}v d^{n}xdt\nonumber\\
    &=&\int^{T}_{0}\int_{\Sigma}\left(u_{tt}v\right)d^{n}xdt+\int^{T}_{0}\int_{\Sigma}\gamma^{ij}\gamma u_{j}v_{i} d^{n}xdt\nonumber\\\nonumber
&=&\int^{T}_{0}<u_{tt},v>dt+\int^{T}_{0}B[u,v;t]dt \\\nonumber
\end{eqnarray}

The final equation is the definition of a weak solution provided the
regularity of the solution and the metric allows the integral to be
well defined. This is indeed the case given the geometric conditions
\ref{con}. The Sobolev embedding theorem in one dimension implies that
$u\in C([0,T], L^{2}(\Sigma))\cap C^{1}([0,T], H^{-1}(\Sigma))$ and
therefore condition \ref{weak2} makes sense. 
%conditions imply continuity, no need to ask for it.just boundness

\subsection{Main result}

The main result we prove can be formally
stated as follows

\begin{Theorem} \label{t1} {\bf Well-posedness in $H^1$} Let
  $(\Sigma_{(0,T]},g_{ab})$ be a
 region of a spacetime satisfying
  the geometric conditions \ref{con}. Then the region $\Sigma_{(0,T]}$
  is wave-regular. That is the wave equation is well-posed in the
  following sense: Given $(u_{0},h)\in H^{1}_{0}(\Sigma_{0})\times
  L^{2}(\Sigma_{0})$ there exists a unique 
weak solution $u$ in $H^{1}(\Sigma_{(0,T]})$ of \ 
  $\square_{g}u=f$ in the sense of definition \ref{def} with initial
  conditions 
\begin{enumerate}
   \item  $u(0,\cdot)=u|_{\Sigma_{0}}=u_{0}$  
  \item $u_{t}(0,\cdot)=\frac{\partial u}{\partial t}|_{\Sigma_{0}}=h$
  \end{enumerate}
  that is stable with respect to initial data in 
  $H^{1}_{0}(\Sigma_{0})\times L^{2}(\Sigma_{0})$.  Moreover, under the hypothesis of Lemma \ref{l2} the
  components of the energy momentum tensor associated to the solution
  satisfy $T^{ab}[u]\in C^{0}([0,T], L^{1}(\Sigma))$.
 
\end{Theorem}

\subsection{Existence of solutions}
 
 To prove existence of
solutions we employ {\emph{Galerkin's method}}. This requires several
steps. We begin by showing uniqueness and existence of approximate
solutions, then we establish a uniform estimate for the solutions and
finally we take a limit in a proper weak topology which converges
to the required weak solution.

We start by choosing smooth functions $w_{k}(x)$ such that:
 
$$\{w_{k}\}^{\infty}_{k=1} \mbox{ is an orthogonal basis of } H^{1}_{0}(\Sigma)$$

$$\{w_{k}\}^{\infty}_{k=1} \mbox{ is an orthonormal basis of } L^{2}(\Sigma)$$

We can form the desired basis by choosing the eigenvectors of the Laplace operator $\Delta$ in the given local coordinates \cite{evans}.

Now fix a positive integer $m$, write

\begin{equation}
\label{defu}
u^{m}(t,x):=\sum^{m}_{k=1}d^{k}_{m}(t)w_{k}(x)
\end{equation}

and consider for each $k=1,...,m$ the equation:

\begin{equation}\label{aprox}
 \left(u^{m}_{tt}, w_{k}\right)_{L^{2}(\Sigma)}+B[u^{m},w_{k};t]=\left(f,w_{k}\right)_{L^{2}(\Sigma,N\nu_{\gamma})}
\end{equation}

where $\nu_{\gamma}$ is the volume form associated to the induced metric $\gamma_{ij}$ on $\Sigma$.

 The system of equations (\ref{aprox}) can be arranged as a system of
linear ODE's given by
 
\begin{equation}
\ddot{d}^{k}_{m}(t)+\sum^{m}_{l=1}e^{kl}(t)d^{l}_{m}(t)=f^{k}(t)
\end{equation}

where $e^{kl}(t):=B[w_{l},w_{k};t]=\int_{\Sigma}\gamma^{ij}(t,x){\gamma(t,x)} w_{l_{i}}w_{k_{j}} dx^{n}$ and $f^{k}(t):=(f,w_{k})_{L^{2}(\Sigma,N\nu_{\gamma})}$ for each $k=1,...,m$ .

We also require that the system satisfies the initial conditions
\begin{equation}\label{aproxic}
 d^{k}_{m}(0)=(u_{0},w_{k})_{L^{2}(\Sigma_{0})},\quad\dot{d}^{k}_{m}(0)=(h,w_{k})_{ L^{2}(\Sigma_{0})}
\end{equation}
for $k=1,...,m$. 

The functions $e^{kl}(t)$ are continuous in $t$ as $\gamma^{ij}(t,x){\gamma(t,x)}\in C^{1}([0,T], L^{\infty}(\Sigma))$. Then by standard local existence and uniqueness theorems for linear ordinary differential equations we obtain a unique $d^{k}_{m}(t)\in C^{2}([0,T])$ for every $k=1,...,m$. 

Therefore we have shown that for each $m$ there is a unique solution, $u^{m}$,  satisfying (\ref{aprox}) and (\ref{aproxic}) which we call the \emph{m-approximate solution}.

\subsubsection{Energy estimates}

In this section we establish the following energy estimate.

\begin{Theorem}
There exists a constant C, depending only on $\Sigma,T$ and the coefficients of $L$ such that
\begin{equation}
\begin{split}\label{eec2as}
  max_{t\in(0,T]}\left(||u^{m}(t,\cdot)||_{H^{1}_{0}(\Sigma)}+||u^{m}_{t}(t,\cdot)||_{L^{2}(\Sigma)}+||u^{m}_{tt}||_{L^{2}([0,T];H^{-1}(\Sigma))}\right)\\
      \le C\left(||f||_{L^{2}([0,T];L^{2}(\Sigma,N\nu_{\gamma}))}+||u_{0}||_{H^{1}_{0}(\Sigma)}+||h||_{L^{2}(\Sigma)}\right)
      \end{split}
\end{equation}

\end{Theorem}

We start by multiplying equality (\ref{aprox}) by $\dot{d}^{k}_{m}(t)$, sum from $k=1,...,m$ and use (\ref{defu})
to obtain

\begin{equation}\label{e0}
  \left(u^{m}_{tt}, u^{m}_{t}\right)_{L^{2}(\Sigma)}+B[u^{m},u^{m}_{t};t]=\left(f,u^{m}_{t}\right)_{L^{2}(\Sigma,N\nu_{\gamma})}
\end{equation}

Using the fact that  that

\begin{equation}\label{e1}
   \left(u^{m}_{tt},u^{m}_{t}\right)_{L^{2}(\Sigma)}=\frac{1}{2}\frac{d}{dt}||u^{m}_{t}||_{L^{2}(\Sigma)}^{2}
\end{equation}

and that

\begin{equation}\label{e}
 B[u^{m},u^{m}_{t};t]=\frac{d}{dt}\left(\frac{1}{2}B[u^{m},u^{m};t]\right)-\frac{1}{2}\int_{\Sigma}\left(\gamma^{ij}(t,x)\gamma(t,x)\right)_{t} u^{m}_{i}u^{m}_{j}
\end{equation}

we have

\begin{equation}\label{e2}
 B[u^{m},u^{m}_{t};t]\ge \frac{d}{dt}\left(\frac{1}{2}B[u^{m},u^{m};t]\right)- C_{1}||u^{m}||^{2}_{H^{1}_{0}(\Sigma)}
\end{equation}

Combining equations (\ref{e0}), (\ref{e1}), (\ref{e}) and (\ref{e2}) we obtain

\begin{eqnarray}
 && \frac{d}{dt}\left(||u^{m}_{t}||_{L^{2}(\Sigma)}^{2}+B[u^{m},u^{m};t]\right) \\
 &\le& C_{2}\left(||u^{m}_{t}||_{L^{2}(\Sigma)}^{2}+||u^{m}||_{H^{1}_{0}(\Sigma)}^{2}+||f||_{L^{2}(\Sigma,N\nu_{\gamma})}^{2} \right)\\ \label{1e3}
  &\le& C_{3}\left(||u^{m}_{t}||_{L^{2}(\Sigma)}^{2}+B[u^{m},u^{m};t]+||f||_{L^{2}(\Sigma,N\nu_{\gamma})}^{2} \right) 
  \end{eqnarray}
  
  Where we have applied the uniform ellipticity condition in order to use the inequality
  
  \begin{equation}\label{ue}
    \theta\int_{\Sigma}|\delta^{ij}u^{m}_{i}u^{m}_{j}|\le B[u^{m},u^{m};t]
  \end{equation}

If we now define the "energy" $E(t)$ of a solution by:

\begin{equation}
  E(t)=||u^{m}_{t}(t,\cdot)||_{L^{2}(\Sigma)}^{2}+B[u^{m},u^{m};t]
\end{equation}

Then inequality (\ref{1e3}) reads

\begin{equation}
  \frac{d}{dt}E(t)\le C_{3} E(t)+C_{3}||f(t,\cdot)||_{L^{2}(\Sigma,N\nu_{\gamma})}^{2} 
\end{equation}

and an application of Gronwall's inequality gives the estimate

\begin{equation}
  E(t)\le e^{C_{4}t}\left(E(0)+C_{3}\int^{t}_{0}||f(t,\cdot)||_{L^{2}(\Sigma,N\nu_{\gamma})}^{2} \right)
\end{equation}

However, we also have 

\begin{equation}
 E(0)\le C_{5}\left( ||u_{0}||_{H_{0}^{1}(\Sigma)}^{2}+||h||_{L^{2}(\Sigma)}^{2}\right) 
\end{equation}
which follows from the initial conditions for the approximate solutions together with $||u^{m}(0)||_{H_{0}^{1}(\Sigma)}^{2}\le||u_{0}||_{H_{0}^{1}(\Sigma)}^{2}$, $||u_{t}^{m}(0)||_{L^{2}(\Sigma)}^{2}\le||h||_{L^{2}(\Sigma)}^{2}$.

Thus we obtain

\begin{equation}\label{1eec2}
\begin{split}
  max_{t\in(0,T]}\left(||u^{m}_{t}||_{L^{2}(\Sigma)}^{2}+B[u^{m},u^{m};t]\right)\\
      \le C_{6}\left(||f||^{2}_{L^{2}([0,T];L^{2}(\Sigma,N\nu_{\gamma}))}+||u_{0}||^{2}_{H^{1}_{0}(\Sigma)}+||h||^{2}_{L^{2}(\Sigma)}\right)
  \end{split}
\end{equation}

Now we have from equation (\ref{aprox}) that

  \begin{equation}
  \begin{split}
 \left(u^{m}_{tt},w_{k}\right)_{L^{2}(\Sigma)}=-B[u^{m},w_{k};t]+\left(f,w_{k}\right)_{L^{2}(\Sigma,N\nu_{\gamma})}\\
 \le C_{7} \left(||u^{m}||_{H^{1}_{0}(\Sigma)}+ ||f||^{2}_{L^{2}(\Sigma)} \right)||w_{k}||_{H^{1}_{0}(\Sigma)}
 \end{split}
\end{equation}

where we have used the bounds on $N,\sqrt{\gamma}$ given by the geometric 
condition 1 and the Cauchy-Schwartz inequality.

Since $(u^{m}_{tt},w_{k})=0$ for $k>m$ by construction we have proved that

\begin{eqnarray}
\displaystyle
  ||u^{m}_{tt}||_{H^{-1}(\Sigma)}&=& \sup_{v\in span\{w_{k}\}}\frac{(u^{m}_{tt},v)_{L^{2}(\Sigma)}}{||v||_{H^{1}_{0}(\Sigma)}}\\
 &\le&C_{8} \sup_{v\in span\{w_{k}\}}\frac{|(f,v)_{L^{2}(\Sigma)}|+|B(u^{m},v;t)|}{||v||_{H^{1}_{0}(\Sigma)}} \\
  &\le& C_{9}\left(||f||_{L^{2}(\Sigma))}+||u^{m}||_{H^{1}_{0}(\Sigma)}\right)
\end{eqnarray}

Squaring the above inequality, integrating in time and using  equation (\ref{1eec2}) we obtain

\begin{equation}
 \int_{0}^{T} ||u_{tt}||^{2}_{H^{-1}(\Sigma)}dt\le C_{12}\left(||f||^{2}_{L^{2}([0,T];L^{2}(\Sigma,N\nu_{\gamma}))}+||u_{0}||^{2}_{H^{1}_{0}(\Sigma)}+||h||^{2}_{L^{2}(\Sigma)}\right)
\end{equation}

which concludes the proof. $\boxdot $

\subsubsection{Convergence to solutions}\label{prov}

We have shown that $\{u^{m}\}^{\infty}_{m=1}$ is bounded in $L^{2}(0,T;H^{1}_{0}(\Sigma))$, $\{u_{t}^{m}\}^{\infty}_{m=1}$ is bounded in $L^{2}(0,T;L^{2}(\Sigma))$ and $\{u^{m}_{tt}\}^{\infty}_{m=1}$ is bounded in $L^{2}(0,T;H^{-1}(\Sigma))$

We now make use of the following Theorem \cite{evans} 
\begin{Theorem}
Let  $X$ be a reflexive Banach space and suppose the sequence $\{u_{k}\}^{\infty}_{k=1}\subset X$ is bounded. Then there exist a sub-sequence $\{u_{k_{j}}\}^{\infty}_{j=1}\subset \{u_{k}\}^{\infty}_{k=1}$ and $u\in X$ such that $u_{k_{j}}\rightharpoonup u$, i.e., $\{u_{k_{j}}\}$ converges weakly to $u$.  
\end{Theorem}

Using the theorem there exists a sub-sequence of approximate functions $\{u^{m_{l}}\}^{\infty}_{l=1}$ such that

\begin{itemize}
  \item  $u^{m_{l}}\rightharpoonup L^{2}(0,T;H^{1}_{0}(\Sigma))$
  \item  $u_{t}^{m_{l}}\rightharpoonup L^{2}(0,T;L^{2}(\Sigma))$
  \item  $u_{tt}^{m_{l}}\rightharpoonup L^{2}(0,T;H^{-1}(\Sigma))$
  \end{itemize}
 
 In order to show this is a weak solution we must
  now verify that the limit of the sequence satisfies conditions 1 and
  2 of definition \ref{def}.

To verify condition 1 we multiply (\ref{aprox}) by a function $\phi(t)\in
  C^{\infty}([0,T])$ and integrate with respect to time to give
  
\begin{equation}\label{aprox0}
\begin{split}
\int_{0}^{T}\left( \left(u^{m_{l}}_{tt}, \phi(t)w_{k}\right)_{L^{2}(\Sigma)}+B[u^{m_{l}},\phi(t)w_{k};t]\right)dt=\int_{0}^{T}\left(f,\phi(t)w_{k}\right)_{L^{2}(\Sigma,N\nu_{\gamma})}dt\\
\end{split}
\end{equation}

Then taking the limit as $m_{l}\rightarrow \infty$ we obtain

\begin{equation}\label{aprox1}
\begin{split}
\int_{0}^{T}\left( < u_{tt},\phi(t)w_{k}>+B[u,\phi(t)w_{k};t]\right)dt=\int_{0}^{T}\left(f,\phi(t)w_{k}\right)_{L^{2}(\Sigma,N\nu_{\gamma})}dt
\end{split}
\end{equation}

%This follows from the fact that the sequence $\{\frac{\sqrt(\gamma)}{N} u^{m_{l}}_{tt}\}^{\infty}_{l=1}$ is bounded in $L^{2}(0,T;H^{-1}(\Sigma))$ and converges to $\frac{\sqrt(\gamma)}{N} u_{tt}$, the fact that the function $\partial_{t}\frac{\sqrt{\gamma}}{N}\phi(t)w_{k}\in L^{2}(0,T;L^{2}(\Sigma))$ and the definition of weak convergence.

Thus for any test function of the form  $v=\sum^{N}_{k=1}\phi^{k}(t)w_{k}(x)$ we have that equality (\ref{aprox1}) is satisfied. Moreover, test functions of that form are dense in $L^{2}(0,T;H^{1}_{0}(\Sigma))$. Therefore, we have shown that 

\begin{equation}\label{aprox2}
\begin{split}
\int_{0}^{T}\left( <u_{tt},v>+B[u,v;t]\right)dt=\int_{0}^{T}\left(f,v\right)_{L^{2}(\Sigma,N\nu_{\gamma})}dt=\int_{\Sigma_{(0,T]}}fv\mu_{g}
\end{split}
\end{equation}

for any $v\in L^{2}(0,T;H^{1}_{0}(\Sigma))$. $\boxdot $

Finally we need to verify that the solution also satisfies the initial conditions. Using the initial condition (\ref{aproxic}) and the fact that $\{w_{k}\}$ is a basis of $L^{2}(\Sigma_{0})$ we obtain

\begin{eqnarray}
  u^{m_{l}}(0,\cdot)&\rightarrow& u_{0} \mbox{ in }L^{2}(\Sigma_{0})\\
  u_{t}^{m_{l}}(0,\cdot)&\rightarrow& h \mbox{ in } L^{2}(\Sigma_{0})
  \end{eqnarray}

Then using the fact that $L^{2}$ convergence of a sequence of functions to a function in $L^{2}$ implies there is a sub-sequence that converges a.e. pointwise we can find a sub-sequence $\{ u^{m_{l_{r}}}(0,\cdot)\}^{\infty}_{r=1}$ such that it converges a.e. pointwise to $u_{0}$. Notice that then we can take a sub-sequence $\{ u^{m_{l_{r_{s}}}}(0,\cdot)\}^{\infty}_{s=1}$ of the sub-sequence $\{ u^{m_{l_{r}}}(0,\cdot)\}^{\infty}_{r=1}$ to add the a.e. pointwise convergence of the sequence $\{ u_{t}^{m_{l_{r_{s}}}}(0,\cdot)\}^{\infty}_{s=1}$ to the function $h$.  $\boxdot $

Therefore, the limit $u^{m_{l_{r_{s}}}}\rightharpoonup u$ gives the desired weak solution.

\subsection{Uniqueness and stability with respect to the initial data}

The proof of uniqueness and stability relies on the energy estimate (\ref{eec2as}). By letting $m$ tend to infinity and using the fact that the norm is sequentially weakly lower-semicontinuous \cite{evans}, we obtain the bound that the weak solution satisfies

\begin{equation}\label{eec2}
\begin{split}
  max_{t\in(0,T]}\left(||u(t,\cdot)||_{H^{1}_{0}(\Sigma)}+||u_{t}(t,\cdot)||_{L^{2}(\Sigma)}\right)\\
        \le C\left(||f||_{L^{2}([0,T];L^{2}(\Sigma,N\nu_{\gamma}))}+||u_{0}||_{H^{1}_{0}(\Sigma)}+||h||_{L^{2}(\Sigma)}\right)
  \end{split}
\end{equation}

Therefore, if $u=u_{1}-u_{2}$ is the difference between two weak solutions satisfying the same initial conditions $u_{0},h$ with the same source function $f$, then $u$ is a weak solution with vanishing initial data $u_{0}=h=0$ and source function $f=0$. 

Hence 

\begin{equation}
  \left(||u(t,\cdot)||_{H^{1}_{0}(\Sigma)}+||u_{t}(t,\cdot)||_{L^{2}(\Sigma)}\right)\le 0
\end{equation}
for all $0\le t\le T$ which implies $u=0$ and therefore $u_{1}=u_{2}$.

We now prove the stability of the solution with respect to
initial data. To make the concept precise we say that the
solution $u$ is continuously stable in $H^{1}({\Sigma}_{(0,T]})$ with respect to initial data in $H^{1}_{0}(\Sigma_{0})\times L^{2}(\Sigma_{0}) $,  if given $\epsilon >0$ there is a $\delta$
depending on $u_{0},h,f$ such that if:
\begin{equation}
  \Arrowvert u_{0}-\tilde{u_{0}}\Arrowvert_{H^{1}_{0}(\Sigma_{0})}\le \delta,
\hbox{ and }
  \Arrowvert h-\tilde{h}\Arrowvert_{L^{2}(\Sigma_{0})}\le \delta
\end{equation}
for $(u_{0},h)\in H^{1}_{0}(\Sigma_{0})\times L^{2}(\Sigma_{0}) $ and 
\begin{equation}
  \Arrowvert f-\tilde{f}\Arrowvert_{L^{2}([0,T];L^{2}(\Sigma,N\nu_{\gamma}))}\le \delta
\end{equation}

for $f\in_{L^{2}([0,T];L^{2}(\Sigma,N\nu_{\gamma}))}$
then
\begin{equation}
  \Arrowvert u-\tilde{u}\Arrowvert_{H^{1}_{0}({\Sigma}_{(0,T]})}\le \epsilon
\end{equation} 

where $\tilde{u}$ is a solution with initial data given by
$\tilde{u}|_{\Sigma_{0}}=\tilde{u_{0}}$ and
$\tilde{u}_{t}|_{\Sigma_{0}}=\tilde{h}$
with source function $\tilde{f}$.

Now squaring (\ref{eec2}) and integrating in time from $[0,\tau\le T]$ we have:
\begin{eqnarray}
  \left(\Arrowvert u-\tilde{u}\Arrowvert_{{\Sigma}_{(0,T]}}\right)^{2}&\le& \int_{0}^{\tau}\left(||u(t,\cdot)||_{H^{1}_{0}(\Sigma)}^{2}+||u_{t}(t,\cdot)||^{2}_{L^{2}(\Sigma)}\right) dt\\ \nonumber
  &\le& K  \int_{0}^{\tau}\left(||u_{0}-\tilde{u_{0}}||^{2}_{H^{1}_{0}(\Sigma)}+||h-\tilde{h}||^{2}_{L^{2}(\Sigma)}\right)+ \Arrowvert f-\tilde{f}\Arrowvert_{L^{2}([0,T];L^{2}(\Sigma,N\nu_{\gamma}))}dt
\end{eqnarray}

for $\tau\le T$ and with $K$ a suitable constant.

Now choosing $\delta=\frac{\epsilon}{\sqrt{3 \tau C}}$ we obtain the inequality:

\begin{equation}
   \left(\Arrowvert u-\tilde{u}\Arrowvert_{{\Sigma}_{(0,T]}}\right)^{2}
\leq K \int_{0}^{\tau}\frac{2\epsilon^{2}}{3 \tau K}+ \frac{\epsilon^{2}}{3 \tau K} dt \leq \epsilon^{2}
\end{equation}
which establishes stability with respect to the initial data.

\subsection{Integrability of the energy momentum tensor} 

The regularity of the solutions allows us to make sense of the energy
momentum tensor of the scalar field $u$ as a tensor with
$L^{1}(\Sigma_{(0,T]})$ components given by
 
\begin{equation}
       T^{ab}[u] = \left(g^{ac}g^{bd}-\frac{1}{2}g^{ab}g^{cd}\right)u_{c}u_{d}-\frac{1}{2}g^{ab}u^{2}\label{energytensor}
\end{equation}

We now show that $T^{ab}[u](t,\cdot)$ is in $L^{1}(\Sigma_{t})$ for
all $0\le t\le T$. To prove this, notice that it is enough to show
that $u\in C([0,T];H^{1}_{0}(\Sigma))\cap
C^{1}(0,T;L^{2}(\Sigma))$. This result also allows us to establish the
existence and uniqueness of solutions in $\Sigma_{(0,T]}$ given
initial data on any hypersurface $\Sigma_{t}$ with $0\le t\le T$. In
this section we closely follow the exposition given in \cite{hunter}.

\begin{Proposition}\label{continuity}
  Let $u$ be a weak solution as defined in Definition \ref{weakcs}
  with $f\in L^{2}(\Sigma_{(0,T]})$ and initial data $(u_{0},h)\in
  H^{1}_{0}(\Sigma_{0})\times L^{2}(\Sigma_{0})$. Additionally let the
  metric satisfies the geometric condition \ref{con} and the hypothesis of Lemma \ref{l2}. Then $u\in
  C([0,T];H^{1}_{0}(\Sigma))\cap C^{1}(0,T;L^{2}(\Sigma)) $
\end{Proposition}

To prove this proposition, we use the following Lemmas

\begin{Lemma}\label{cw}
Suppose that $V, H$ are Hilbert spaces and $V\hookrightarrow H$ is densely and
continuously embedded in $H$. If
\begin{equation}
  u \in L^{\infty} (0, T ; V) ,
u_{t} \in L^{2} (0, T ; H) ,
\end{equation}

then $u \in C_{w} ([0, T ]; V)$ is weakly continuous.

\end {Lemma}

Then from the fact that $H^{1}_{0}\hookrightarrow L^{2}\hookrightarrow H^{-1}$ and the energy estimate we have that $u\in C_{w}([0,T];H^{1}_{0}(\Sigma))$ and $u_{t}\in C_{w}([0,T];L^{2}(\Sigma))$ 

\begin{Lemma}\label{l2}
Let $u$ be a weak solution that satisfies $u_{tt}+L u\in L^{2}(0,T; L^{2}(\Sigma))$ and assume there is a mollified sequence of functions given by  
\begin{equation*}
  \{u^{\epsilon}(t,x)=\eta^{\epsilon}(t)*(\xi u(t,x))\}
\end{equation*}

where $\{\eta^{\epsilon}(t)\}_{\epsilon}$ is a strict delta net and
$\xi$ is a smooth cut-off function vanishing outside the interval
$(0,T)$ and equal to one in some sub-interval $I\subset (0,T)$
such that $u^{\epsilon}_{tt}+L u^{\epsilon}\in L^{2}(0,T; L^{2}(\Sigma))$ and in the limit $\epsilon\rightarrow 0$ we have  $u^{\epsilon}_{tt}+L u^{\epsilon}\rightarrow u_{tt}+L u$ in $ L^{2}(0,T; L^{2}(\Sigma))$

Then

\begin{equation}\label{ew}
   \frac{1}{2}\frac{d}{dt}\left(||u_{t}||_{L^{2}(\Sigma)}^{2}+B[u,u;t]\right)=(u_{tt}+Lu,u_{t})_{L^{2}(\Sigma)}^{2}+\frac{1}{2}B_{t}[u,u;t]
\end{equation}

where
\begin{equation}
  B_{t}[u,v;t]:=\int_{\Sigma}\sum^{n}_{i,j=1}\left(\gamma^{ij}(t,x)N(t,x)\sqrt{\gamma(t,x)}\right)_{t} u_{x^{i}}v_{x^{j}} dx^{n}
\end{equation}
and
\begin{equation}\label{ec}
    E(t)=\left(||u_{t}(t,\cdot)||_{L^{2}(\Sigma)}^{2}+B[u,u;t]\right):(0,T]\rightarrow \mathbb{R}
    \end{equation} 
  is an absolutely continuous function.

\end{Lemma}

\emph{Proof.}

We have that

\begin{eqnarray}\nonumber
  &\frac{1}{2}\frac{d}{dt}\left(||u^{\epsilon}_{t}||_{L^{2}(\Sigma)}+B[u^{\epsilon},u^{\epsilon};t]\right)\\ \nonumber
  =&(u_{tt}^{\epsilon},u^{\epsilon}_{t})_{L^{2}(\Sigma)}^{2}+B[u^{\epsilon},u_{t}^{\epsilon};t]+\frac{1}{2}B_{t}[u^{\epsilon},u^{\epsilon};t] \\ \label{e3}
 =& (u_{tt}^{\epsilon}+Lu^{\epsilon},u^{\epsilon}_{t})_{L^{2}(\Sigma)}+\frac{1}{2}B_{t}[u^{\epsilon},u^{\epsilon};t]
  \end{eqnarray}

   Taking the limit as $\epsilon\rightarrow 0$ give us the same result for the unmollified function $\xi u$ and hence equation (\ref{ew}) holds on every compact sub-interval of
   $(0,T)$.

 Now the derivative of the RHS of equation
   (\ref{ec}) is in $L^{1}(0,T)$ since using equation (\ref{e3}) and taking the limit $\epsilon\rightarrow 0$ we
   have 
 
\begin{eqnarray}\nonumber
  \int_{0}^{T}\left(u_{tt}+Lu,u_{t}\right)_{L^{2}(\Sigma)}dt&\le&||u_{tt}+Lu||_{L^{2}([0,T];L^{2}(\Sigma))}||u_{t}||_{L^{2}([0,T];L^{2}(\Sigma))}\\ \nonumber
  &\le&||f||_{L^{2}([0,T];L^{2}(\Sigma))}||u||_{L^{2}([0,T];L^{2}(\Sigma))}
\end{eqnarray}

\begin{eqnarray}
  \int_{0}^{T}B_{t}[u,u;t]&\le& C ||u||^{2}_{L^{2}([0,T];H^{1}_{0}(\Sigma))}
\end{eqnarray}

Thus $E(t)$ is the integral of a $L^{1}$ function, so is absolutely continuous. $\boxdot$ 

We can now prove Proposition \ref{continuity}.\\

{\emph{Proof of Proposition \ref{continuity}.}}

Using the weak continuity of $u_{t}$, the continuity of $E$ and the continuity of $a_{t}$ in $H^{1}_{0}$ we find that

\begin{equation}
\begin{split}\nonumber
||u_{t}(t,\cdot)-u_{t}(t_{0},\cdot)||_{L^{2}(\Sigma)}^{2}+B[u(t,\cdot)-u(t_{0},\cdot),u(t,\cdot)-u(t_{0},\cdot);t_{0}]\\
 =||u_{t}(t,\cdot)||_{L^{2}(\Sigma)}^{2}+||u_{t}(t_{0},\cdot)||_{L^{2}(\Sigma)}^{2}\\+B[u(t,\cdot),u(t,\cdot);t_{0}]+B[u(t_{0},\cdot),u(t_{0},\cdot);t_{0}]\\-2(u_{t}(t,\cdot),u_{t}(t_{0},\cdot))_{L^{2}(\Sigma)}-2B[u(t,\cdot),u(t_{0},\cdot);t_{0}]\\
 =||u_{t}(t,\cdot)||_{L^{2}(\Sigma)}^{2}+B[u(t,\cdot),u(t,\cdot);t] \\+||u_{t}(t_{0},\cdot)||_{L^{2}(\Sigma)}^{2}+B[u(t_{0},\cdot),u(t_{0},\cdot);t_{0}]\\-2B[u(t,\cdot),u(t_{0},\cdot);t_{0}]-2(u_{t}(t,\cdot),u_{t}(t_{0},\cdot))_{L^{2}(\Sigma)}\\ +B[u(t,\cdot),u(t,\cdot);t_{0}]-B[u(t,\cdot),u(t,\cdot);t]\\
 =E(t)+E(t_{0})+B[u(t,\cdot),u(t,\cdot);t_{0}]-B[u(t,\cdot),u(t,\cdot);t]\\-2\left((u_{t}(t,\cdot),u_{t}(t_{0},\cdot))_{L^{2}(\Sigma)}+B[u(t,\cdot),u(t_{0},\cdot);t_{0}]) \right)
  \end{split}
\end{equation}

Therefore taking the limit $t\rightarrow t_{0}$ we obtain 

\begin{eqnarray}\label{cee}
\begin{split}
& \displaystyle\lim_{t\rightarrow t_{0}} \left(||u_{t}(t,\cdot)-u_{t}(t_{0},\cdot)||_{L^{2}(\Sigma)}^{2}+B[u(t,\cdot)-u(t_{0},\cdot),u(t,\cdot)-u(t_{0},\cdot);t_{0}]\right)\\
  &=\displaystyle\lim_{t\rightarrow t_{0}} E(t)+E(t_{0})+B[u(t,\cdot),u(t,\cdot);t_{0}]-B[u(t,\cdot),u(t,\cdot);t]\\&-2\left((u_{t}(t,\cdot),u_{t}(t_{0},\cdot))_{L^{2}(\Sigma)}+B[u(t,\cdot),u(t_{0},\cdot);t_{0}]) \right)\\
 &= E(t_{0})+E(t_{0})-2\left((u_{t}(t_{0},\cdot),u_{t}(t_{0},\cdot)+B[u(t_{0},\cdot),u(t_{0},\cdot);t_{0}])\right)\\
&=E(t_{0})+E(t_{0})-2\left(||u_{t}(t_{0},\cdot)||_{L^{2}(\Sigma)}^{2}+B[u(t_{0},\cdot),u(t_{0},\cdot);t_{0}])\right)=0
 \end{split}
\end{eqnarray}

Then using equation (\ref{ue}) we have

\begin{eqnarray}
  &\displaystyle\lim_{t\rightarrow t_{0}}\left(||u_{t}(t,\cdot)-u_{t}(t_{0},\cdot)||_{L^{2}(\Sigma)}^{2}+\theta||u(t,\cdot)-u(t_{0},\cdot)||^{2}_{H^{1}_{0}(\Sigma)}\right)\\ \nonumber
  &\le\displaystyle\lim_{t\rightarrow t_{0}} \left(||u_{t}(t,\cdot)-u_{t}(t_{0},\cdot)||_{L^{2}(\Sigma)}^{2}+B[u(t,\cdot)-u(t_{0},\cdot),u(t,\cdot)-u(t_{0},\cdot);t_{0}]\right) 
\end{eqnarray}

Hence using (\ref{cee}) we conclude

\begin{equation}
\begin{split}
  \lim_{t\rightarrow t_{0}}||u_{t}(t,\cdot)-u_{t}(t_{0},\cdot)||_{L^{2}(\Sigma)}=0\\
  \lim_{t\rightarrow t_{0}}||u(t,\cdot)-u(t_{0},\cdot)||_{H^{1}_{0}(\Sigma)}=0
  \end{split}
  \end{equation}
So that $u$ is an element of $C([0,T];H^{1}_{0}(\Sigma))\cap C^{1}(0,T;L^{2}(\Sigma)) $ as required.
$\boxdot$ 

\end{section}
\begin{section}{Discussion}\label{Discussion}
  In this section we discuss how Theorem \ref{t1} applies to several
  physical scenarios. We treat the case of spacetimes with cosmic
  strings and show that these spacetimes, despite having regions where
  the curvature behaves as a distribution, or in the case of dynamic
  cosmic strings can even develop curvature singularities, are
  $H^1$-wave regular. We end by discussing how the concept of
  $H^1$-wave regularity is related to the Strong Cosmic Censorship Conjecture.  
  
\subsection{Spacetimes with Cosmic Strings}

Cosmic strings are topological defects that potentially formed during
a phase transition in the early universe. Current observations put
tight constrains on the dimensionless string tension $G\mu\le 10^{-8}$
in Planck units where $c=1$, $G=m_{pl}^{-2}$ and $\mu$ is the mass per
length \cite{Ligo}. Additionally, the effective thickness of a cosmic string is
of the order $10^{-29}$ cm. \cite{mathem}. This extremely small width justifies
what is called "the thin string limit". This is the metric around a
static infinitely straight Nambu-Goto string lying along the z-axis
satisfying Einstein's equations, which is "conical" in the plane
transverse to the string with the line element given by 
\begin{equation}
ds^{2}=dt^{2}-dz^{2}-d\rho^{2}-(1-4G\mu)^{2}\rho^{2}d\theta^{2} 
\end{equation}
where  $0\le\theta<2\pi$.

By introducing a new angular coordinate
$\tilde{\theta}=(1-4G\mu)\theta$, the spacetime can be seen to be flat
everywhere except at $\rho=0$ where there is an angular deficit of
$2\pi(1-A)$ with $A=(1-4G\mu)$. We want to consider the region
containing $\rho=0$ so we transform to Cartesian coordinates $(x,y)=(\rho \cos\theta,\rho
\sin\theta)$ which are regular at $\rho=0$ and rewrite the line element as
 
\begin{equation}
  ds^{2}=dt^{2}-\frac{x^{2}+A^{2}y^{2}}{x^{2}+y^{2}}dx^{2}-\frac{2xy(1-A^{2})}{x^{2}+y^{2}}dxdy-\frac{y^{2}+A^{2}x^{2}}{x^{2}+y^{2}}dy^{2}-dz^{2}
\end{equation}
Notice that the metric has a direction dependant limit on the $z$-axis so fails to be $C^{0}$ at the axis although it remains bounded.

By direct inspection one can see that the metric is bounded
everywhere, $N=1$, $\sqrt{\gamma}=A$ and given that $0<A\le 1$ the
uniform ellipticity condition is satisfied. These conditions imply
that under a rescaling of the time coordinate the hypothesis of
Theorem \ref{t1} is satisfied.  Therefore, this spacetime is
$H^{1}$-wave regular.

In fact, we can consider a time dependent generalisation of this metric given
\begin{equation}
\begin{split}
  ds^{2}=dt^{2}-\frac{x^{2}+A^{2}(t)y^{2}}{x^{2}+y^{2}}dx^{2}-\frac{2xy(1-A^{2}(t))}{x^{2}+y^{2}}dxdy\\-\frac{y^{2}+A^{2}(t)x^{2}}{x^{2}+y^{2}}dy^{2}-dz^{2}
\end{split}
\end{equation}

This spacetime represents a dynamical cosmic string with a time
dependent deficit angle. The metric satisfies the conditions of the
Theorem \ref{t1} as long as the angle deficit satisfies $0<A(t)\le1$
and the function $A(t)$ is $C^{1}$. Moreover, the dynamical cone can
develop curvature singularities in contrast with the static cone
\cite{dynamic}.

 Vickers \cite{cs3,cosmic} showed that under certain conditions
two-dimensional quasi-regular singularities can be seen as
generalised strings. Moreover, the strings are
totally geodesic and only the normal directions to the string are
degenerate. For timelike generalised cosmic strings this guarantees
that the time derivatives are not problematic. 

% In fact, the conservation of holonomy which implies the totally geodesic condition requires that the lapse function is $C^{2}$ and $N\rightarrow 1$ when $\rho\rightarrow 0$. Therefore, without loss of generality we can choose $N=1,\beta^{i}=0$.

Finally, assuming that that the generalised cosmic string admits a
$3+1$ splitting given by a family of $L^{\infty}$ Riemaniann metrics
with suitable lapse and shift (see geometric conditions \ref{con}) and
the angle deficit is chosen such that the uniform ellipticity
condition is satisfied, we can conclude that generalised cosmic
strings are $H^{1}$-wave regular.

Notice however that the spinning cosmic string metric given by
\begin{equation}
  ds^{2}=(dt^{2}+4Jd\theta)^{2}-dr^{2}-A^{2}r^{2}d\theta-dz^{2}
\end{equation}
does not satisfy the hypothesis of the theorem since $\beta^{i}\neq0$ and there are no local coordinates containing a neighbourhood of the $z$-axis which make $\beta^i$ vanish.
\newpage
 
 \subsection{The Strong Cosmic Censorship Conjecture}

 As mentioned in the introduction, the singularity theorems only
 establish geodesic incompleteness and not the reason for the
 incompleteness.  In examples such as the Kerr and
 Reissner–Nordstr\"om spacetimes there is a loss of global
 hyperbolicity instead of a loss of regularity. Mathematically a
 spacetime region ${\mathcal N}$ is said to globally hyperbolic if the
 causality condition is satisfied and for any two points $p, q \in
 {\mathcal N}$ the causal diamond $J^+(p) \cap J^-(q)$ is compact and
 contained in ${\mathcal N}$ \cite{bernal}.  One demands this
 condition because it is sufficient to guarantee the \emph{global}
 well-posedness of the wave equation and other physical fields
 \cite{bar}. Therefore, a spacetime that fails to be globally
 hyperbolic signals the possibility of a loss of predictability of the
 initial value problem of any field on it including the metric.  While
 the maximal globally hyperbolic development is unique \cite{hawking}
 one may extend the spacetime (even in a $C^{\infty}$ manner) in
 non-unique ways if one does not demand global hyperbolicity. The
 boundary of the original manifold in the extension is known as the
 Cauchy horizon. It is expected that under small perturbations the
 extensions are unstable and some loss of regularity will occur
 preventing the extension. This is part of the content of The Strong
 Cosmic Censorship Conjecture. In fact, Dafermos has shown that the
 outcome of the conjecture depends critically on the differentiability
of the metric  allowed in the extensions of the maximal Cauchy
 development \cite{daf2}.
 
As we noted in the
introduction and showed in our results, a necessary condition for a singularity
to be regarded as a strong gravitational singularity is that in any neighbourhood of it the evolution of the wave equation is not well posed. In this spirit, a well-motivated physical
formulation of predictability may include the criteria of locally
well-posedness of test field as a necessary condition. We therefore propose the following condition for ``predictability''. 
 
\begin{Condition}
  If an initial value set for Einstein's equations is said to be
  predictable then there is no nontrivial future extension $(M,g)$ of
  its maximal domain of development such that the extension is
  $H^{1}$-wave regular.
\end{Condition}

 \end{section}
\section*{Acknowledgements}

The authors would like to thank CONACyT for supporting this work through a CONACyT Graduate Fellowship.

\end{document}